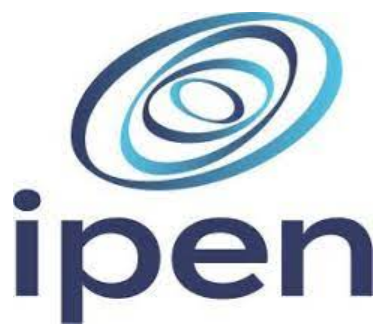

**AUTARQUIA ASSOCIADA À UNIVERSIDADE DE SÃO PAULO**

Long-term BVOCs fluxes in the Amazon rainforest by Relaxed Eddy Accumulation coupled to Gas Chromatograph Photoionization Detector (REAPER)

**CAROLINE KAKO OSTERMANN**

**Qualification presented as part of the requirements to obtain the Degree of Doctor of Science in the Area of Nuclear Technology - Materials.**
**Advisor: Dr. José Oscar William Vega Bustillos**
**Co-advisor: Dr. Alex B. Guenther**

São Paulo
2023

# SUMMARY

# 1 INTRODUCTION

The annual global emission of all reactive carbon is estimated to be about 2.4 Pg (1 Pg=$10^{15}$ g) of carbon (Muller et al., 2016; Guenther et al; 1992). These estimates include approximately 0.6 Pg C as carbon monoxide (CO), 0.5 Pg C as methane, 0.5 Pg C as isoprene, and 0.8 Pg C as other volatile organic compounds (VOC). About half of this total is associated with emissions from terrestrial vegetation; the remainder is primarily from technological sources, biomass burning, and microbes. (Guenther et al., 2002).

Isoprene ($C_5H_8$) dominates the emission of biogenic volatile organic compounds (BVOCs) into the atmosphere, and its major global source is tropical vegetation (Guenther et al., 2012). Emitted by vegetation, it has been linked to temperature regulation, reducing drought-induced stress and other physiological processes within plants (Sharkey et al., 2008; Sharkey et al., 2013).

A dialkene, isoprene is prone to oxidation by reaction with the hydroxyl radical (OH) as well as by ozonolysis and reaction with the nitrate radical ($NO_3$) (Stone et al., 2011). Isoprene oxidation pathways are complex (Archibald et al., 2010) and result not only in several oxygenated volatile organic compounds (OVOCs *e.g.,* formaldehyde, methacrolein and methyl-vinyl ketone) but also in a suite of low-volatility stable products and intermediates that can act as precursors of secondary organic aerosols (SOA) (Carlton et al., 2009; Clayes et al., 2004; Liu et al., 2016).

As a result of its high reactivity and large emissions, determining the global abundance of isoprene is important to understand the oxidizing capacity of the atmosphere (Alves et al., 2018) and the formation of SOA, which can affect the optical properties of the atmosphere and, in turn, impact the climate (Squire et al., 2015).

With its high plant foliage biomass and rich plant diversity (ter Steege et al., 2013), the Amazon rainforest represents a key source of isoprene to the atmosphere. However, model estimates of isoprene emission and its intra- and interannual variability in the Amazon still carry high uncertainty because only a few observational experiments have been conducted with mechanistic and process-based approaches, which hinders further modeling optimization (Yanez-Serrano et al., 2015).

Some reasons for uncertainties in isoprene model estimates are already known. The correct determination of the magnitude of the isoprene source – or the emission factor at leaf standard conditions (1000 µmolm$^{-2}$s$^{-1}$ photosynthetically active radiation – PAR, 30°C), as it is conceptualized in models (Muller et al., 1992; Guenther et al., 1995) – is crucial to improve isoprene modeling estimates. The Amazonian plant biodiversity represents a considerable challenge for determining the isoprene emission factor. Although previous studies suggested that ~1% of tree species are hyperdominant – with their tree individuals responsible for half of all tree stems, carbon storage, and productivity (ter Steege et al., 2013; Fauset et al., 2015).

One of the most critical knowledge gaps is how the tropical forest isoprene fluxes differs under extremely hot and dry conditions and the season transitions, such as in El Niño years, and how this might affect atmospheric processes. As some studies have indicated that extreme years will become more frequent and intense with climate change (Nobre et al., 2016) it is essential to understand the processes mediated by isoprene in such years to improve model estimates (Yanez-Serrano et al., 2020).

The unique example of the most pristine ecosystems on the Earth is the Amazon Forest, located in the northern part of South America and comprises an area of 6.8 million km$^2$ (Wittmann et al., 2016). Approximately 80 % of this region is covered with rainforest, which accounts for circa 40 % of all tropical forests on the globe (Goulding et al., 2003; Andrade et al., 2015; Leppla et al., 2023).

Measurements of atmospheric hydrocarbons such as isoprene are challenged by the difficulty in making measurements in remote places like in tropical forests. To date, in situ measurements of isoprene have been carried out using existing commercial bench-top instruments, such as gas chromatographs (Jones et al., 2011) and mass spectrometers (Fauset et al., 2015).

These techniques differentiate between VOCs either by separation (gas chromatography) or by identification of their molecular ions based on mass-to-charge ratios (mass spectrometry). These instruments, while offering high precision and stability, are not built to withstand field conditions for long periods of time due to their need for power, temperature-controlled environments and specialty carrier gases (Ferracci et al., 2020).

In addition, instrument cost and maintenance time both limit the number of instruments deployed at any one time, and, hence, the spatial coverage of a field campaign. Because of this, there exists a need to fill this missing information with new devices for observatory climate research.

# 2 CHAPTER ONE: Development a Portable, Low-Cost, Energy-Saving *Relaxed Eddy Accumulator Gas Chromatograph Photoionization Detector* (REAPER)

This chapter delves into the creation of an innovative and globally deployable portable device designed to provide extended, real-time measurements of isoprene fluxes. This is achieved through the integration of the Relaxed Eddy Accumulation (REA) (Guenther et al., 1996) methodology with Gas Chromatography Photoionization Detector (GC-PID) (Bolas et al., 2020), ensuring a comprehensive and long-term assessment.

## 2.1 Background and Aspects of Originality

On a global scale, the primary source of Volatile Organic Compounds (VOCs) stems from biogenic processes, as indicated by (Guenther et al., 1995). Consequently, understanding the intricacies of atmospheric chemistry, particularly concerning oxidants and aerosol formation, necessitates a comprehensive assessment of Biogenic VOCs (BVOCs). However, this is a formidable challenge due to the vast diversity of vegetation and ecosystems, along with temporal changes in land use, including land use changes.

The prevalent BVOC emission model was established by (Guenther et al.,1995; Gunther et al., 2012). It categorizes vegetation into broad plant functional types or specific vegetation types, each assigned emission capacities for various BVOC categories. Temporal fluctuations are governed by emission algorithms influenced by observed meteorological variables, primarily radiation and temperature.

Nevertheless, experimental measurement of emission capacities for individual plants to inform these models remains difficult, given the vast array of vegetation types and the additional variability in environmental parameters within natural canopies. This challenge is further compounded by temporal variations in natural systems, such as phenology changes and water stress, which are inadequately represented in global BVOC models (Fu, 2019 ).

Micrometeorological flux techniques offer an observational means of determining BVOC emissions over canopy and land-surface scales, ranging from hectares to a few square kilometers. This facilitates the parameterization of models based on ecosystem type. This is especially valuable in highly diverse ecosystems like tropical forests, where a few hectares can contain hundreds of tree species. Among these techniques, Eddy Covariance (EC) stands out as the most direct and robust method for measuring fluxes. It involves correlating concentration fluctuations with changes in vertical wind velocity, as pioneered by Baldocchi et al. (1988) and Lenschow (1995).

EC requires simultaneous high-frequency measurements (under 1 second) of three-dimensional wind velocities and the concentration of the target species several meters above a suitable vegetation canopy. These measurements must span the entire range of time scales associated with turbulent transport within the atmospheric surface layer. This necessitates fast measurements that can later be subsequently averaged for longer time scales (typically 0.5–1 hour). This requirement sets EC apart from traditional analytical methods, which rely on longer sampling times.

Recent advancements in chemical ionization mass spectrometry have made fast BVOC determinations feasible. However, this technology is non-portable, relatively costly, and demands substantial expertise for field operation. Additionally, it relies on significant power resources, often unavailable at flux tower field sites, and poses maintenance challenges for prolonged studies.

Therefore, there is a pressing need for alternative flux techniques, such as the Relaxed Eddy Accumulator (REA). REA segregates air samples based on rapid (10 Hz) wind fluctuations into different storage reservoirs, which can be analyzed at a later time using slower analytical methods. Since its inception by Businger and Oncley (1990), the REA technique has been adapted for measuring ecosystem-scale fluxes of trace species, encompassing CO2, aerosols, and BVOCs (Greenberg et al., 2003). The complexity of previous REA measurements has ranged from simple to highly intricate.

This project endeavors to introduce an innovative instrumentation system designed to remotely and continuously deliver real-time data on isoprene fluxes (figure 1). This system integrates the REA methodology with a GC-PID for comprehensive and timely measurements. The device is designed to be portable, easy to operate, cost-effective, low-power (AC/DC connection), and lightweight. It is specifically engineered for deployment in challenging environments and remote locations without mains power (figure 2). This device not only ensures accurate VOC chromatograms but also addresses the limitations of previous methods, offering improved temporal resolution and reduced resource requirements.

**Figure 1:** The components of the REAPER system, each categorized as follows: (a) REA box, (a.1) Inlet, (a.2) Sonic anemometer; (b) Inert bags box; (c) GC-PID, (c.1) Isoprene standard, (c.2) Nitrogen ($N_2$); (d) Power plug adaptor box; (e) Outlet box, (e.1) Satellite antenna, (e.2) Internet modem box.

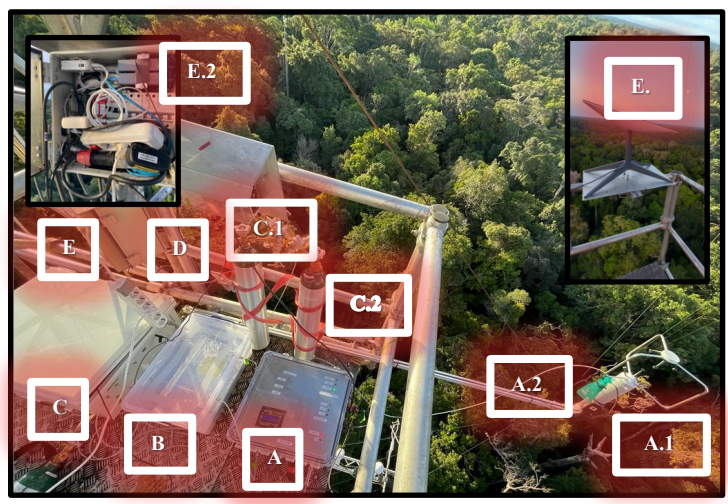

**Source:** From the author, 2023.

**Figure 2:** Field coordinates (-2.136639, -59.006306) at the tower the REAPER was deployed.

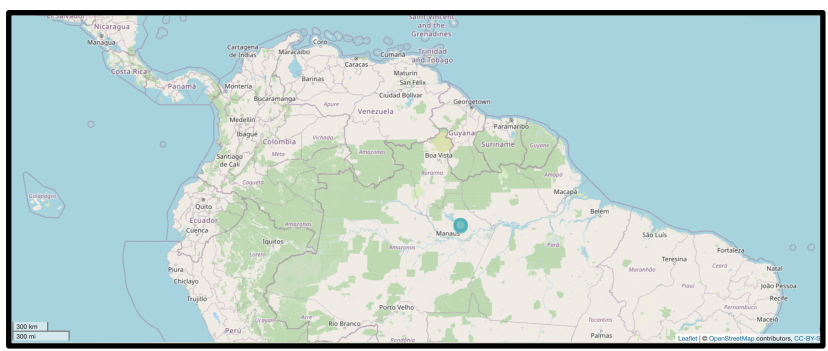

**Source:** From Ostermann et al., 2024. (Under construction).

While recent work has demonstrated the feasibility of retrieving isoprene abundances in the boundary layer using satellite measurements based on thermal infrared imaging (Fu, 2019), these retrievals entail uncertainties ranging from 10% to 50% and would benefit from validation by ground-based instrumentation.

In addition to its measurement capabilities, the novel instrument can transmit data remotely via a cloud-based platform, such as Google Drive, providing real-time access to meteorological data, calculated parameters, chromatograms, and flux results. This feature enables the deployment of these systems across a wide range of sites, with the objective of obtaining BVOC flux measurements across existing flux tower networks, thereby assessing the magnitude of emitted compounds and estimating their emission capacities across various ecosystems.

## 2.2 Objectives

The primary goal of this project was to develop, evaluate and deploy a groundbreaking device designed for remote field applications. This endeavor can be explained through the following key steps:

*a.) System Design, Schematic Blueprint and Software Development:*

- At the BAILAB under the supervision of Dr. Alex Guenther at University of California, Irvine (UCI/United States).

*b.) Environmental Testing and Data Memory.*
*c.) GC-PID Calibration and Peak Analysis*
*d.) Deployment in Remote Environments:*

- The deployment of the device it was at Amazon Tall Tower Observatory site (ATTO) with a collaboration of Dr. Jonathan Williams from max Planck Institute for Chemistry (MPIC/Germany). A team from Dr. Sergio Duvoisin at University of Amazon Estate (UEA/Brazil) was trained to operate and maintain the instrument during the field campaign.

*e.) Analysis of Flux Data:*

- At the Engineering Science Laboratory of Harvard University (United States), under the guidance of Dr. Scot Martin

## 2.3 Methodology

The idea of this research was to conceptualize, develop, and assess an instrument founded upon the REA methodology, enabling the real-time qualification and quantification of BVOCs fluxes on a global scale. This investigation primarily targets remote locations, addressing the notable absence of empirical data within existing global climate models, and potentially serving as an inaugural step toward the creation of novel climate models.

### 2.3.1 Approach Overview

The REAPER is a novel sampling apparatus engineered for the purpose of segregating atmospheric air according to the instantaneous vertical wind velocity components. This complex system facilitates the collection and segregation of air samples into two distinct inert storage reservoirs: one designated for upward-moving air currents and the other for downward-moving air masses.

The allocation of air samples into these reservoirs is contingent upon the prevailing direction of the vertical wind velocity at the precise moment of sample acquisition. This instrument has been designed to be cost-effective, energy efficient, and structurally robust.

At its core, the REAPER system (Figure 3), is tailored to channel air samples into an online GC-PID (Figure 4), representing its principal role in the sampling of Isoprene. This compound, renowned for its prominence in forested environments, represents the primary target compound for analysis in conjunction with VOCs.

**Figure 3:** REAPER air flow diagram.

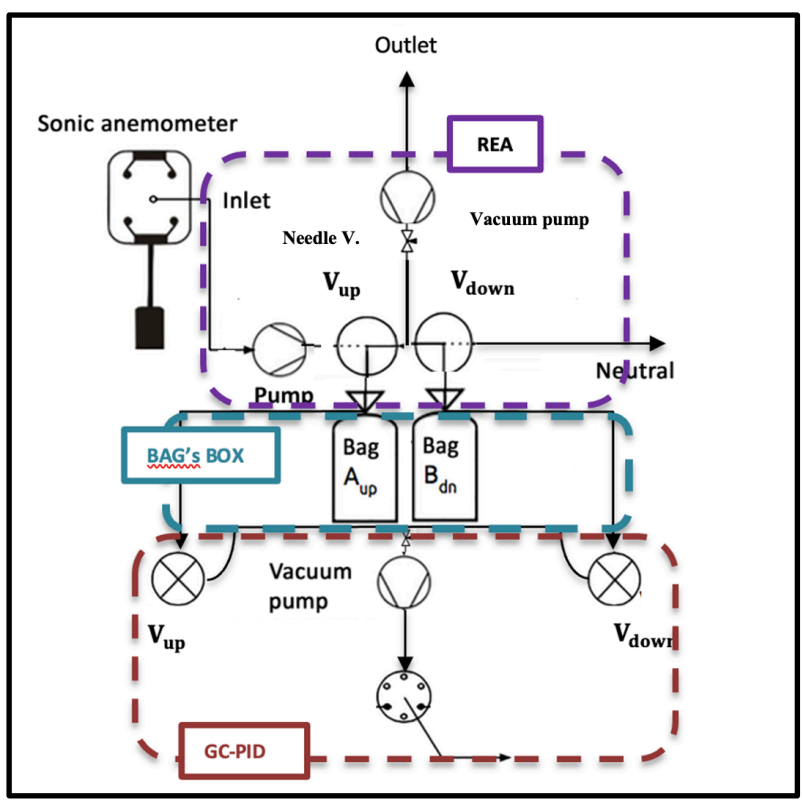

**Source:** From the author, 2023.

**Figure 4:** Schematic GC-PID diagram.

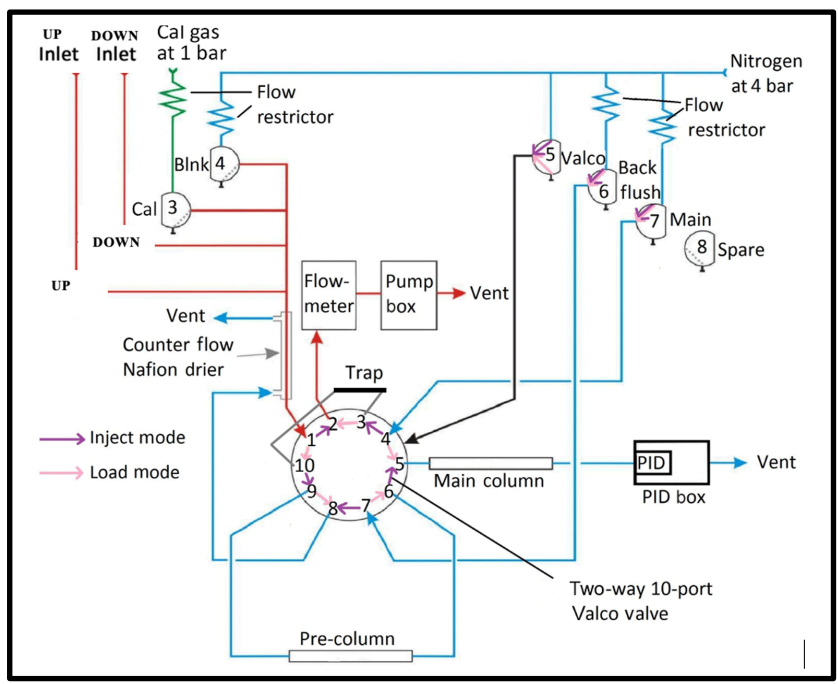

**Source:** Adapted from (Bolas et al., 2020).

Additionally, this versatile device offers the capability to generate a spreadsheet of outputs hosted on Google Drive platform during its operation. This spreadsheet encompasses data points denoting averages and standard deviations derived from meteorological parameters such as wind components (U, V, and W), temperature, and heat flux. Additionally, the REAPER is equipped to calculate the β-coefficient (see Equation 2) and document data from the anemometer.

This functionality of this data dissemination is facilitated through the utilization of either a Raspberry Pi board or an Ethernet connection, seamlessly interfacing with the Arduino MEGA board. Furthermore, it communicates autonomously and concurrently with the GC-PID via a serial cable connection.

A tangible manifestation of the REAPER in operation is depicted in the accompanying figure 1, which captures a moment during the Amazon rainforest campaign at the ATTO site. This specific installation is situated at "INSTANT" tower, standing at an elevation of 58 meters height, and was deployed in May 2023.

The REAPER Version 1.0 comprises three integral components, namely: *(a)* a Main Control unit housing the power supply infrastructure and governing datalogger, *(a.1)* the Segregator enclosure encompassing the sampling valves along with a flowmeter and *(a.2)* a Sonic Anemometer of the RM Young brand, specifically the Model 81000, *(b)* An enclosure, constructed with inert material bags, serves as a receptacle for the segregated air samples, distinguished by their vertical direction, that are to be gathered, *(c)* the GC unit a trap column is incorporated, followed by a secondary column containing *Dimethyl TBS Cyclodextrin* (Ostermann, 2022; Bolas et al., 2020), *(d)* The PID unit *(e)* The Main Control unit fulfills a pivotal role, encompassing the provision of power infrastructure.

### 2. 3.1.1 Operating Principles and Flux Calculations

The follow essential components:

1) The R.M. Young Company's Ultrasonic Anemometer Model 81000.
2) Controller: acquire winds, control segregator (MEGA-Arduino and Raspberry Pi board, signal cables), SD card and cloud communication.
3) Segregator: direct air flow (valves).
4) Flow system: move sample air (pump, tubing, inlet with filter, flowmeter).
5) Reservoirs (Two 6L inert bags): store sample air until it can be analyzed.
6) Power system (AC/DC): battery, charger, power cable bi volt (110/220V).
7) Gas Chromatograph Photoionization Detector (GC-PID) (Bolas et al., 2020).

The fundamental equation (Equation 1) for calculating fluxes ($F_i$) of a specific species from the REA system during this time frame is as follows:

$$F_i = \sigma_w \cdot \beta \cdot (C_{up} - C_{down}) \tag{1}$$

- $F_i$ ($\mu g/(m^2 \cdot s)$).
- $\sigma_w$ denotes the standard deviation of the vertical wind velocity (m/s).
- $\beta$ represents an empirical coefficient (elaborated upon subsequently).
- C_up and C_down signify the concentrations (densities) of the species of interest in the up and down reservoirs, respectively ($\mu g/m^3$).

The $\beta$ coefficient is derived from the sonic temperature and vertical wind velocity fluctuations, obtained by re-arranging the same equation (Equation 2) under the assumption of scalar similarity.

$$\beta = \frac{w'T'}{\sigma_w(T_{up} - T_{down})} \tag{2}$$

- $\beta$ is the beta coefficient, which characterizes the turbulence of the atmosphere.

- $w'$ represents the vertical wind velocity fluctuation (m/s).
- $T'$ represents the temperature fluctuation (K).
- $\sigma_w$ is the standard deviation of vertical wind velocity (m/s).
- $T_{up}$ and $T_{down}$ represent the temperatures in the updrafts and downdrafts, respectively. (K)

The REA box retains all the essential values for flux computation according to (Equation 1 and 2), as well as the heat flux values in (W/m²) (Equation 3).

$$\varphi_q = \rho_{air} \cdot Cp_{air} \times \overline{W'T_{air}'} \tag{3}$$

- $\varphi_q$ represents the latent heat flux in the air (W/m²).
- $\rho_{air}$ is the air density, which represents the air mass/volume (kg/m³).
- $Cp_{air}$ is the specific heat capacity of dry air (J/(kg·°C)).
- $\overline{W'T_{air}'}$ represents the average of the vertical wind velocity and the air temperature covariance (m/s · °C).

The air accumulation in the inert bags is determined through a chromatogram generated by GC-PID box, while concentrations are computed based on the volume and flow that has passed over each respective bag, with volume being ascertained via integration of the flowmeter signal (Bolas et al., 2020)

The REA system necessitates two critical pieces of information at the onset of each flux averaging period:

(1) A mean vertical wind velocity denoted as $\overline{w}$, which is essential for ascertaining the direction of the instantaneous vertical wind velocity (Equation 4).

$$w' = w(t) - \overline{w} \tag{4}$$

- $w'$ represents the vertical wind velocity fluctuation at a specific time (t) (m/s).
- $w(t)$ is the instantaneous vertical wind velocity at that same time (t) (m/s).
- $\overline{w}$ represents the average vertical wind velocity (m/s).

(2) The standard desviation of vertical wind velocity, represented by $\sigma_w$, which one is employed in the calculation of a "deadband."

The deadband comprises a range of small $w'$ values, centered around $\bar{w}$, within which the sampled air is drawn through the "neutral" line. The concentration within this range is inconsequential for flux calculations and is therefore disregarded. The present REA system employs a deadband of $\pm 0.6\sigma_w$. This practice effectively amplifies differences in measured concentrations $(C_{up} - C_{down})$, thereby facilitating the analytical technique employed. It is noteworthy that the $\beta$ coefficient must also be calculated, as per Equation 2, using the same deadband value.

Both $w'$ and $\sigma_w$ are essential at the commencement of a flux averaging period. However, because these values are not available beforehand (constraining the REA technique) they are estimated using the values obtained during the previous flux averaging period, a practice referred to as "memory."

Empirical testing suggests one can use a constant value for both $w'$ and $\sigma_w$, allowing for the consistent utilization of a fixed coordinate reference frame, albeit not a perfectly accurate one.

## 2. 3.1.2 Instrumental Design and Setup

Figure 5 illustrates the fundamental schematic representation of the REA enclosure, encompassing the primary microcontroller, ancillary electronic components, power supply infrastructure, the central air flow control system, and the Raspberry Pi board facilitating data transfer control.

**Figure 5:** Schematic electronical REA diagram.

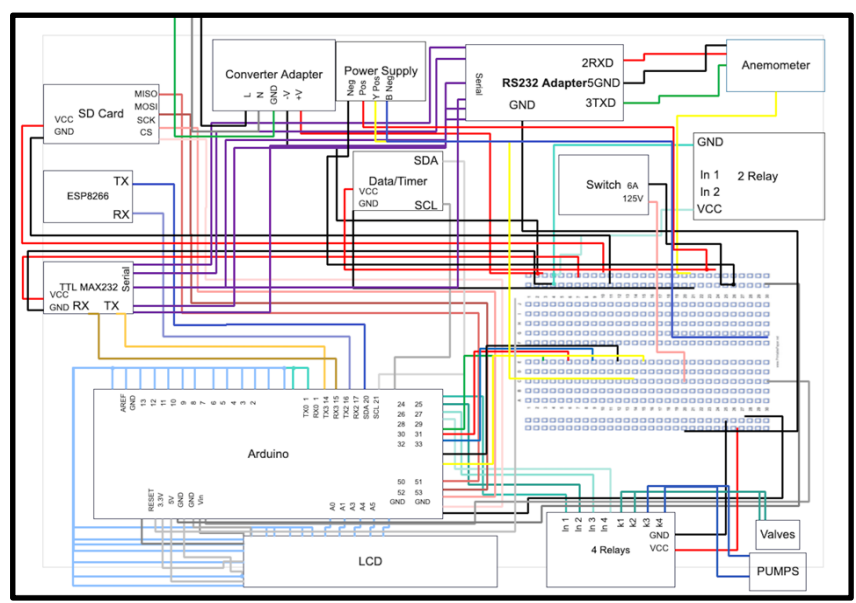

**Source:** From the Author, 2023.

The schematic electronic REAPER diagram was implemented through a comprehensive process. The system relies on Arduino MEGA board programming in the C$^{++}$ language to orchestrate various essential functionalities. These functions encompass monitoring and regulating the flowmeter and anemometer, which involve both analog and digital connections.

Additionally, the Arduino code manages the valves controlling the system and the bags for sample storage, while establishing a serial connection with the GC-PID unit, and the calibration curves of the flowmeter, pump, and anemometer.

To ensure accurate data acquisition and recording, the code also integrates real-time clock (RTC) functionality for precise date and time synchronization. This could be run over a year using less than 0.5 terabytes of storage in the SD card. Furthermore, the REAPER enhances user-friendliness and accessibility by presenting pertinent information on the Liquid Crystal Display (LCD), with the added feature of individual LEDs assigned to each function, and convenient tactile buttons integrated onto the control board (Figure 6).

**Figure 6:** Friendly control board of the REA unit.

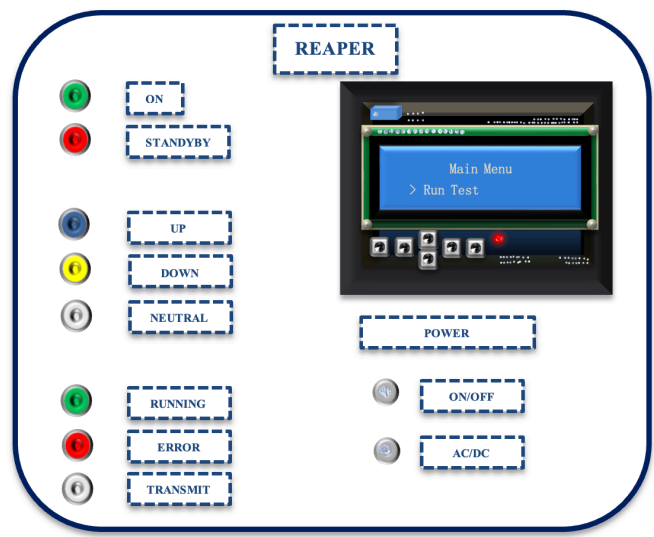

**Source:** From the author, 2023.

The system's operational components, such as switches and relays within the power box, are efficiently controlled through Arduino programming.

Furthermore, implementation includes the incorporation of a Raspberry Pi board programmed in Python to connect to the Google Cloud platform. This connection is instrumental in generating an extensive spreadsheet in 1-kilobyte (.xlsx) format, accommodating a substantial volume of data points without interruptions over the course of a week, encompassing up to 320,000 data points. The Python code is adept at handling data transmission and storage seamlessly.

To enable remote access and control, the system leverages the Virtual Network Computing (VNC) remote connection. The code manages HDMI, Ethernet, and USB 2.0 ports, enhancing the system's versatility and connectivity, all of which contribute to a robust and sophisticated environmental analysis setup with a strong emphasis on data accuracy and reliability.

**2. 3.1.3 GC-PID**

The Gas Chromatography Photoionization Detector apparatus provides voltage values derived from flow and chromatograms, which necessitate calibration and subsequent integration to determine the concentration of isoprene. A procedure to calculate the concentration of a compound using the flow voltage and the voltage of chromatogram peaks is outlined below:

1) Instrument Calibration:
   - Calibrate your chromatography system using known standard solutions of the compound at various concentrations. Run these standards through the system and record the corresponding flow voltages and peak voltages. This data will be used to create a calibration curve.

2) Chromatogram Acquisition:
   - Inject your sample into the chromatograph to obtain a chromatogram. Identify the peaks that correspond to the compound of interest.

3) Peak Integration:
   - Use chromatography software to integrate the area under the peaks. The area under the peak is proportional to the amount of the compound in your sample. You will obtain the peak voltage for each peak.

4) Calibration Curve:

- Plot the calibration curve, which relates the concentration of the compound to the flow voltage or peak voltage. The curve may be linear, quadratic, or follow another mathematical relationship, depending on your system and compound.

5) Calculate Concentration:

- For each peak in your sample, use the peak voltage to find the corresponding concentration on the calibration curve. The equation for calculating concentration from the calibration curve may look like this:

6) Concentration = (Peak Voltage - Intercept)/Slope*
- Peak Voltage: The voltage value of the peak from your sample.
- Intercept: The y-intercept of the calibration curve.
- Slope: The slope of the calibration curve (Slope = ΔC / Δt (or ΔV).

7) Average and Report Results:
- If you have multiple peaks in your chromatogram, calculate the concentration for each peak and average the results if needed. Report the concentrations of the compound in the sample.

## 2.4 Results and Discussions

In the initial phase of REAPER instrument testing, comprehensive evaluations were conducted within the controlled environment of Dr. Alex Gunther's laboratory. This controlled environment was maintained over several weeks to ensure that the system performed in accordance with the anticipated parameters and initiated the generation of the new $\sigma_w$ values.

To verify and validate the system's performance, the heat flux equation was introduced, in line with the methodology proposed by Ostermann (2021). This allowed for a rigorous assessment of the β coefficient's accuracy and reliability.

Data acquisition for the REAPER unit was conducted at a high temporal resolution, with data points being recorded every two seconds. Each cycle, which lasted 30 minutes, marked the calculation of a new *β* coefficient. These data points were collected and synchronized through the Google Drive platform (Figure 7).

Furthermore, to ensure real-time monitoring of flux results, a chromatogram was generated every 50 minutes, depicting voltage magnitude variations. This continuous, non-stop operation ensured that flux results were produced at regular intervals, facilitating a dynamic and responsive monitoring approach.

**Figure 7:** REA spreadsheet from Google Drive cloud platform.

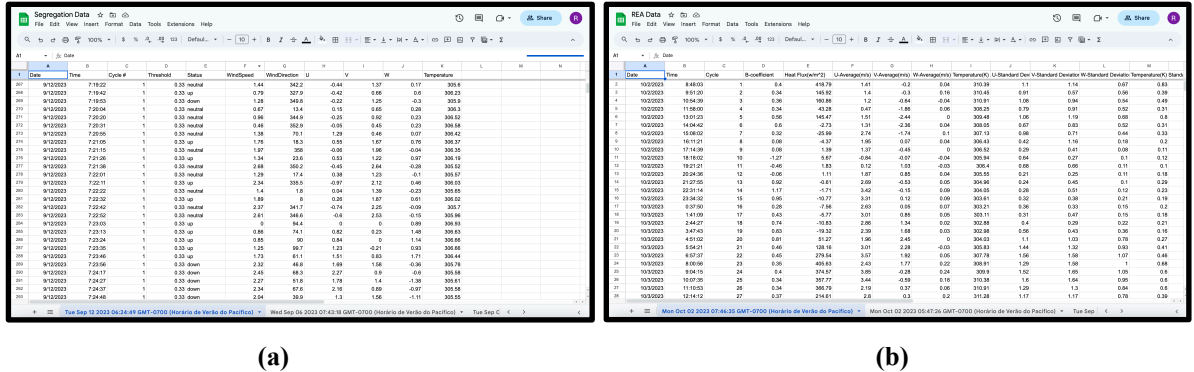

(a)                                                                                          (b)

**Source:** From the author, 2023.

Figure 7 illustrates the organization of data within the REA spreadsheet, hosted on the Google Drive cloud platform. In the first sheet, designated as "Segregation Data" (a), a compilation of meteorological observations is presented. These meteorological data points are recorded at high temporal resolution, with measurements taken at 2-second intervals. This data recording provides a comprehensive history of local meteorological conditions, which is crucial for subsequent analyses and research.

The second sheet, denominated as "REA Data" (b), assumes a central role in this data organization. It serves as the repository for the critical parameters of the study, including the b coefficient and heat flux. Additionally, this sheet features the computation of statistical indicators such as averages and standard deviations for each parameter.

The analysis of the dataset was conducted within the framework of a custom-coded program developed in the RStudio software environment (as outlined by Ostermann et al., 2024, under development). Given the considerable volume of data that required correlation and visualization in graphical form, there is a need to facilitate the portrayal of extensive temporal series involving the β coefficient and its direct correlation with the oscillations in heat flux and temperature.

In the subsequent Figures 8 and 9 (comprising sections a, b, and c) controller environmental at Professor Dr. Alex Gunther's laboratory and in ATTO site consecutively, a notable correlation becomes evident, particularly with respect to the heat flux—both above and below 50 W/m$^2$ - and its clear delineation between diurnal and nocturnal cycles. Additionally, this interrelation is influenced by radiative factors and temperature variations. Notably, the

17range of the β coefficient for the REA system is constrained to fall within the range of 0.30 to 0.45 (Ostermann, 2022), which is considered indicative of the system's proper functioning. Nevertheless, substantial fluctuations, often of high or exceptionally low magnitude, are observed.

These correlations and insights pertaining to the system's operational dynamics hold significant import, particularly when endeavoring to calculate the flux of specific compounds - such as isoprene. It is essential to ensure with a high degree of confidence whether the compound is undergoing deposition or emission during the time of sampling. This is the primary rationale behind designating the system as a "segregator system", underscoring its pivotal role in precise flux determinations within this context.

**Figure 8:** Temporal analyses at REA unit from REAPER control data.

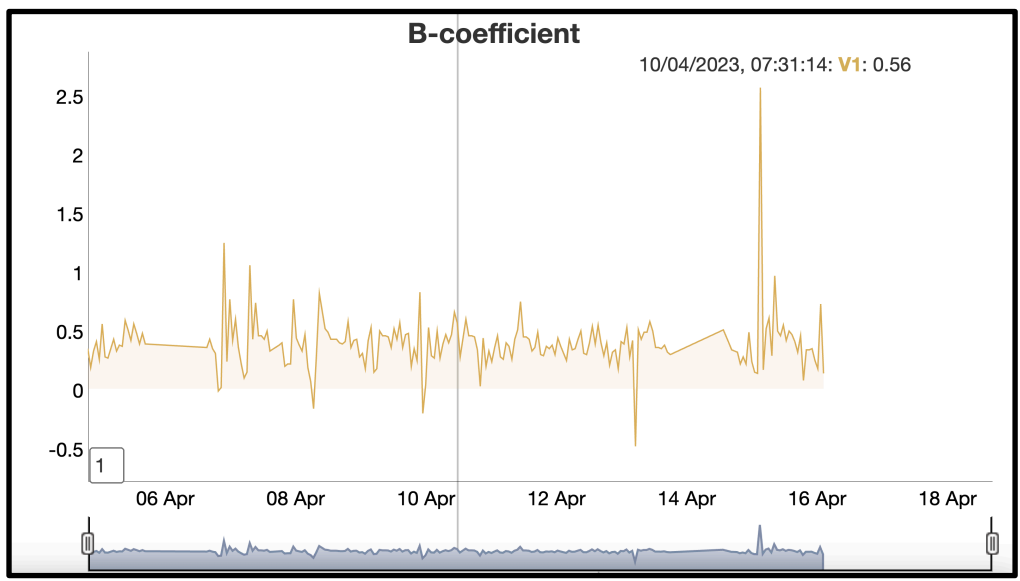

(a)

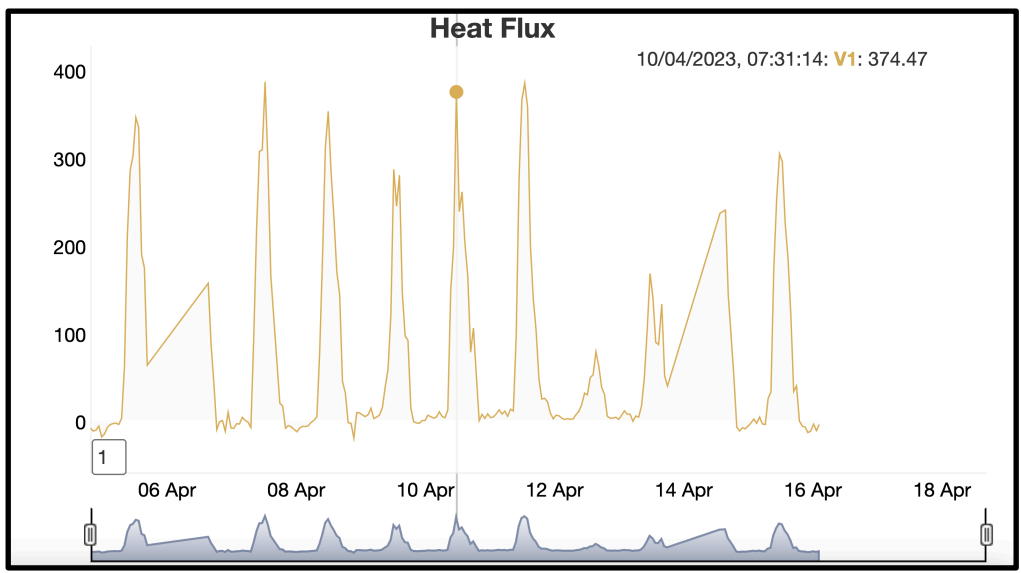

(b)

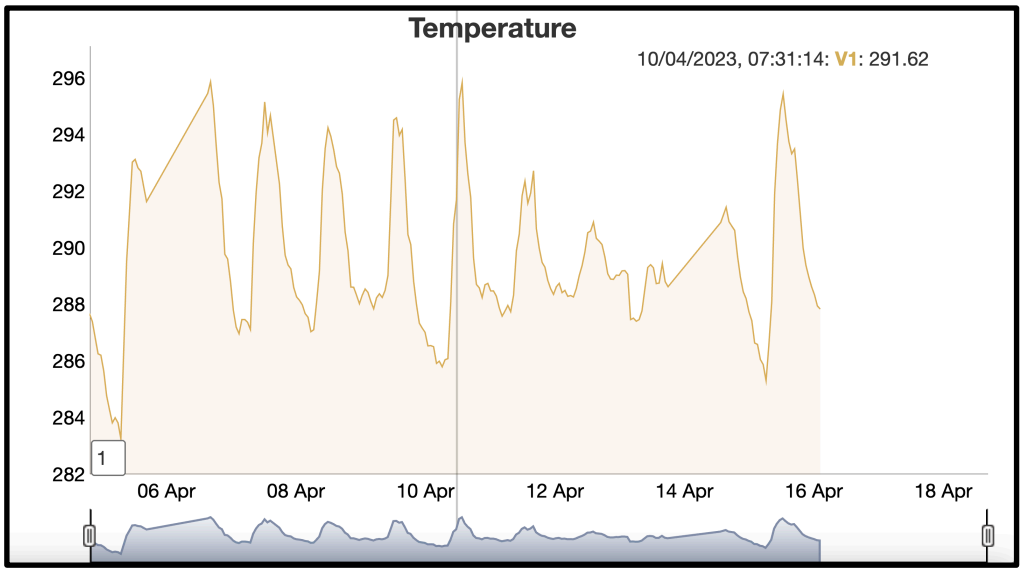

(c)

**Source:** Ostermann et al., 2024 (under construction).

**Figure 9:** Temporal analyses at REA unit from REAPER ATTO site data.

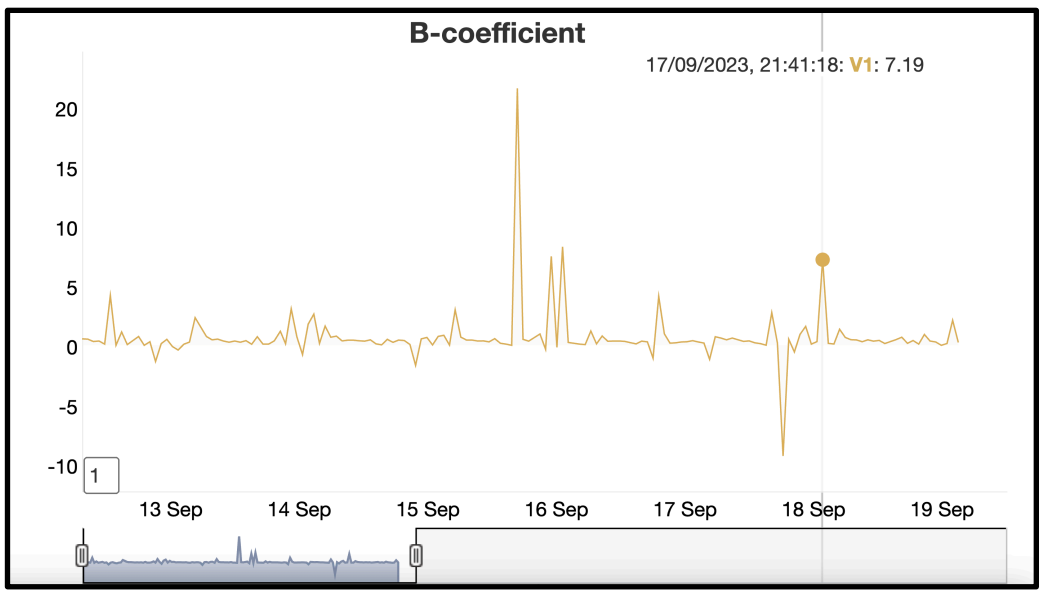

(a)

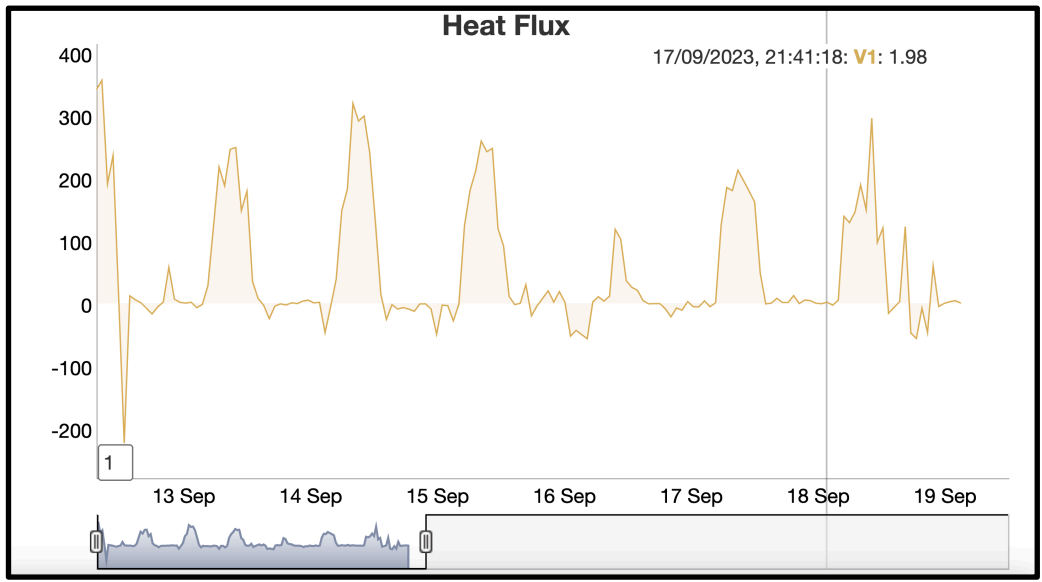

(b)

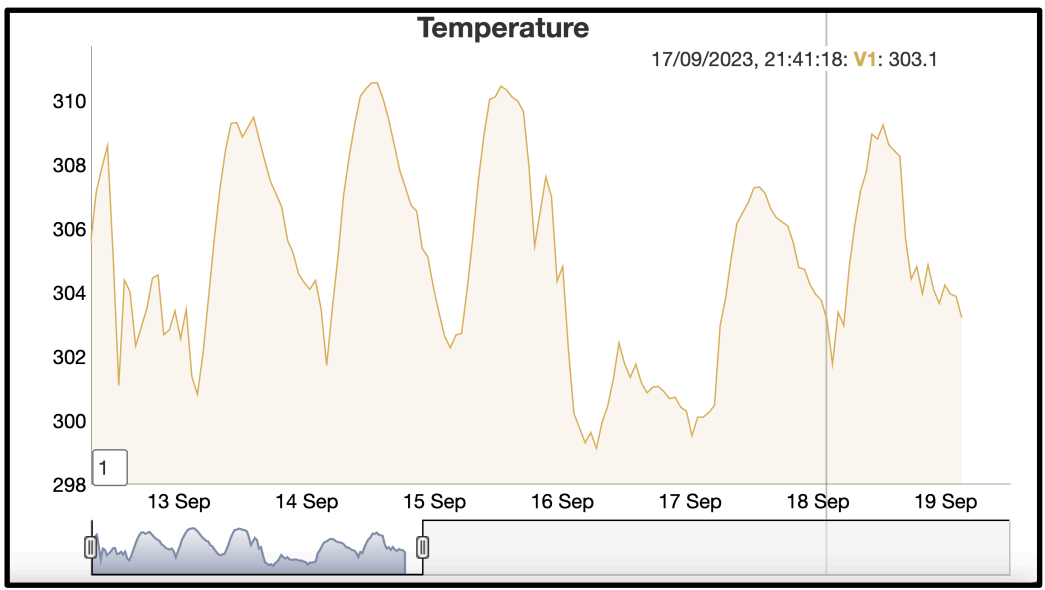

(c)

**Source:** Ostermann et al., 2024 (under construction).

The veracity of these correlations is substantiated by the data depicted in Figure 10 (comprising sections a, b, and c), which illustrates a consistent manifestation of the same effect throughout all the weeks sampled, from September 12[th], 2023 to present, in the Amazon rainforest site. Notably, even during the transitional period from the wet season to the dry season, observed within this timeframe, the established pattern remains conspicuously consistent and unaltered.

**Figure 10:** Temporal analyses at REA unit from REAPER Amazon rainforest data.

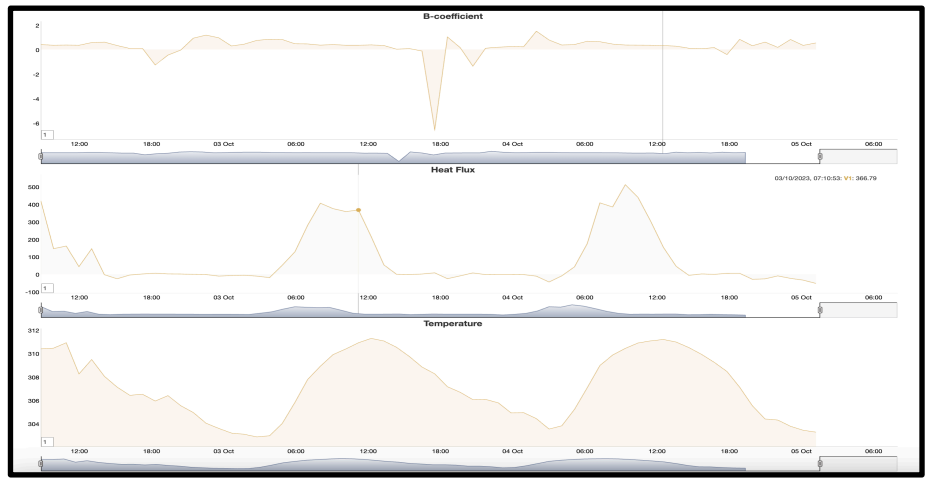

(a)

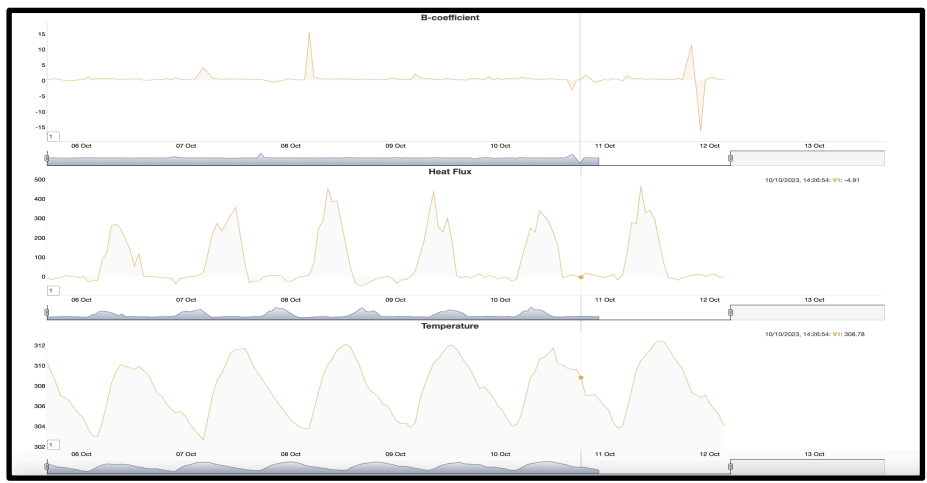

(b)

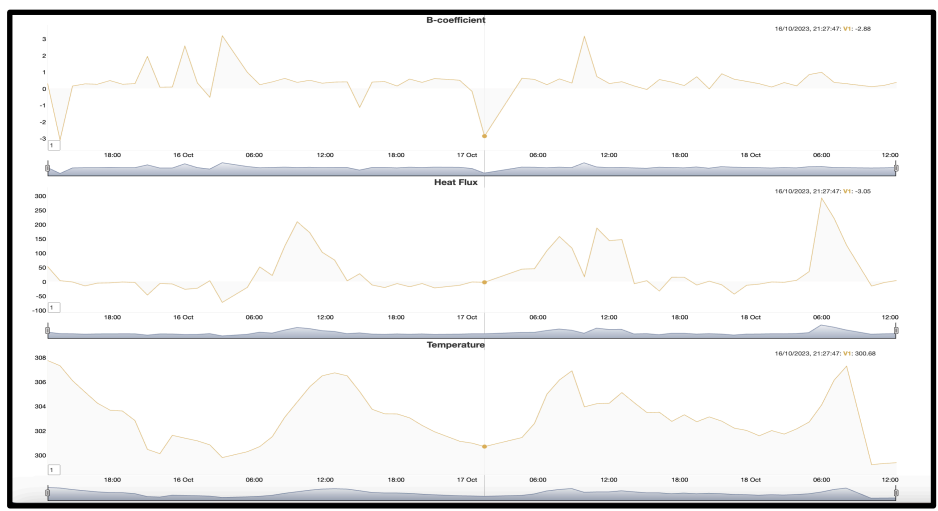

(c)

**Source:** Ostermann et al., 2024 (under construction)

## 3. TIMELINE

| TIME | ACTIVITES |
|---|---|
| Summer 2023 to Fall 2024 | Sampling and data analyses + finish the Chapter 1. |
| Winter 2024 | Development a novel flux model + publish the Chapter 1. |
| Spring 2024 | Write the Chapter 2 (Isoprene Concentrations at Acai+Boat). |
| Summer 2024 to Fall 2025 | Publish the Chapter 2 and Write the Chapter 3 (Isoprene fluxes+modeling). |
| Fall to Winter 2025 | Area Seminar + Publish the Chapter 3. |
| Spring 2025 | Write Thesis. |
| Summer to Fall 2026 | Write Thesis and defend. |